\documentclass[a4paper]{jpconf}
\usepackage{graphicx}
\begin{document}

\title{An experimental apparatus for measuring the Casimir effect at large distances }

\author{P Antonini$^1$, G Bimonte$^2$, G Bressi$^3$, G Carugno$^1$, G Galeazzi$^{4,5}$, G Messineo$^{1,4}$ and G Ruoso$^5$}
\address{$^1$ INFN sez di Padova, via Marzolo 8, 35131 Padova, Italy}
\address{$^2$ Dipartimento di Scienze Fisiche Universit\`a di Napoli Federico II 
Complesso Universitario MSA, Via Cintia, 80126 Napoli, Italy and INFN, Sezione di Napoli, Napoli, Italy}
\address{$^3$ INFN sez. di Pavia, via Bassi 6, 27100 Pavia, Italy}
\address{$^4$ Dipartimento di Fisica, Universit\`{a} di Padova, via Marzolo 8, 35131 Padova, Italy}
\address{$^5$ INFN Lab. Naz. Legnaro, viale dell'Universit\`{a} 2, 35020 Legnaro (Pd), Italy}

\ead{Giuseppe.Ruoso@lnl.infn.it}

\begin{abstract}
An experimental set-up for the measurement of the Casimir
effect at separations larger than a few microns is presented. The
apparatus is based on a mechanical resonator and uses a homodyne
detection technique to sense the Casimir force in the plane-parallel
configuration. First measurements in the 3-10 micrometer range show an unexpected
large force probably due to patch effects.
\end{abstract}



\section{Introduction and motivation}

In quantum electrodynamics the description of the vacuum is modified
by the presence of boundary conditions: the introduction of
conducting surfaces changes the  ground state of the
electromagnetic field in a way that depends on the allowed mode
frequencies. A variation of the position of the surfaces results in a
change of the mode frequencies and thus in an energy change. This
corresponds to a net force acting on the surfaces, and is
known as Casimir effect \cite{casimir48,lifshitz56,bordag01b}. The attractive force
between two parallel and perfectly conducting metal plates is
\begin{equation}\label{eq:casimir1}
   F = \frac{\pi^2 \hbar c S}{240\, d^4},
\end{equation}
where $c$ is the speed of light, $\hbar$ the reduced Planck
constant, $S$ the surface of the plates, and $d$ their separation.

 After the pioneering works of Sparnaay \cite{Sparnaay} and van Blokland and Overbeek \cite{vb}, the experimental study of the Casimir effect has
received a strong impulse in the last decade, following the first
precision measurement conducted by Lamoreaux \cite{lamoreaux97a}.
Several experiments followed these measurements, all but one using a
plane-sphere geometry: Mohideen \textit{et al.} \cite{mohideen98a},
Decca \textit{et al.} \cite{decca03}, Ederth \cite{ederth00} and Chan {\it et al.} \cite {Chan}.
The group of Bressi \textit{et al.}
\cite{bressi02a} was able to perform a measurement in the plane
parallel geometry originally proposed by Casimir. With this geometry
the experimental difficulties are bigger, and this explains why the
precision is worse compared to other configurations. 
More on the subject can be found on this proceedings.

The Casimir
formula (\ref{eq:casimir1}) is valid only in an ideal situation, and
has to be corrected to take into account for example the finite
conductivity of the metal plates and their roughness. The
experimental study of these corrections, that are not negligible only
at very small distances ($\leq$1\,$\mu$m), was possible up to now only
with the plane-sphere configuration due to its better precision.
These measurements were performed for separations smaller than 1
micrometer, while for larger distances the Casimir force was below the
sensitivity of the experiments.

An important correction to the
Casimir formula is due to the presence of thermal photons: the
original calculation has been made at zero temperature but
experiments are normally performed at room temperature. Due to
the form of the black body spectrum, thermal corrections start to be
significative only for separations larger than a few microns,
and become the dominant part for distances larger than
\begin{equation}\label{eq:tempdep}
    \lambda_T = \frac{2 \pi c}{\omega_t} = \frac{\hbar c}{k_B T},
\end{equation}
corresponding to 7~$\mu$m at 300~K \cite{genet00a}.
Since the
separation has to be rather large, the most promising geometry
for observing the thermal contributions to the Casimir force is the
plane parallel one, where the possibility to use relatively wide
surfaces results in  measurable  forces. A plane - cylinder geometry has also been recently
proposed \cite{brown-hayes05,hertzberg05a}.

The effect of thermal radiation in a Casimir like experiment has recently been 
measured by the group of E. A. Cornell \cite{cornell}: in this case a Casimir-Polder configuration has been used \cite{cp}, studying the force between a bulk object and a gas-phase atom \cite{antezza}.

In this paper we present a set-up based on the
plane-parallel geometry where we aim  to measure the
Casimir force for distances of several microns, thus entering in the
regime of strong thermal corrections. Our apparatus is based on an
oscillating plate whose motion is detected using a Michelson
type interferometer. This mechanical resonator is made of aluminized
silicon and its motion is excited at low frequency by means of a
movable source plate. A homodyne detection scheme is employed,
allowing for  very long integration times necessary to measure the small
 motions of the resonator induced by the weak force.
Care has been taken in the design of the apparatus to reduce
background noises and systematic effects. Vibrations are in fact
driving the resonator and are thus limiting the sensitivity and
increasing the integration time. The most important systematic contribution comes
from the presence of voltage biases between the plates even when
they are shortcut. These voltages must be counterbiased with
high precision otherwise they will hide the Casimir force. 

The exact calculation of the finite temperature corrections to the
Casimir force is still an open question: there are at least two
models, both are predicting a sizable change in the force for
separations above 1 - 2 microns. Without the necessity to enter
into this dispute, we quote as an example the results of Bordag \textit{et al.}
\cite{bordag01b}: in the plane-parallel configuration, at a
temperature $T = 300$~K and a separation of  5~$\mu$m, the
correction to the standard force is about 50\%, thus a precision of the
order of 10-20\% in the measurement of the force would then be
sufficient to separate this contribution.

\section{Experimental set-up}\label{sect:exp-set-up}

A detailed description of the apparatus and of the detection technique has been given in 
\cite{nostro}, here only a brief summary will be given.

\subsection{The apparatus}

The plane-parallel configuration is the most convenient geometry for
the measurement of the force at large separations, providing a
possibly measurable signal even at distances of the order of a few
microns. In our set-up two flat parallel surfaces are fronting each
other. One surface is called the source and the other the resonator. The distance between the two
can be changed by means of a piezoelectric actuator which holds the source.
The source can exert a force on the resonator, through an electrostatic field or the
Casimir effect.
The resonator is free to move around its equilibrium position, and by monitoring changes of its
position using a laser interferometer it is possible to  gather
information on the forces acting on it.

The resonator is made of silicon crystal and is essentially a
plate 1 cm $\times$ 1 cm $\times$ 50 $\mu$m connected to a frame
through four parallel wires 1 cm long and 50 $\times$ 50 $\mu$m$^2$ in section.
Plate, wires and frame are carved from the same crystal which was initially
coated with 0.4~$\mu$m of aluminium. We measured
its resonance frequency to be $\nu_r$ = 125 Hz with
a quality factor around 2000.

\begin{figure}[h]
\includegraphics[width=4in]{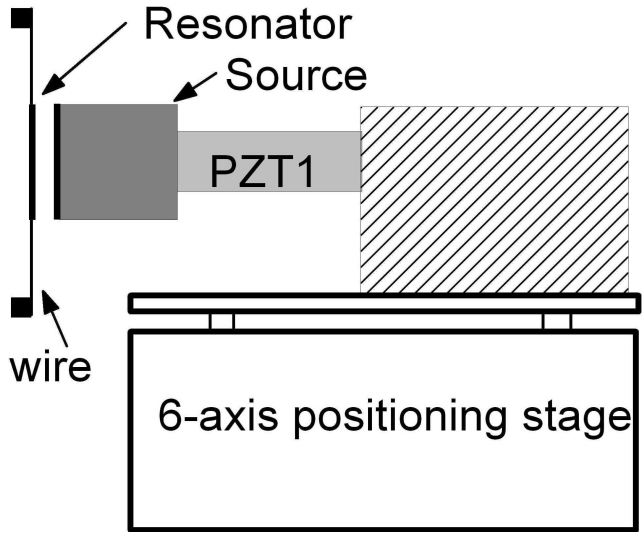}
\caption{Side view showing the resonator, the source and the positioning system used for the parallelization.}
\label{fg:5axis}
\end{figure}

The source is a 1.2 cm $\times$ 1.5 cm, 500 $\mu$m thick aluminium block, with its surface polished
with diamond at the optical level. Measurement of the overall flatness showed a peak to valley distortion less than 200 nm.  The source is
glued to a glass slab, which is in turn glued to a
piezoelectric actuator (PZT1) providing a controllable
translational motion of the source itself.
The PZT is clamped to a 6-axis
stepper-motor-controlled positioning stage (Thorlabs, 6-Axis NanoMax Nanopositioner),
which is used to control the parallelism between the
two plates. This arrangement is shown in figure~\ref{fg:5axis}.

The source, the resonator and the positioning stage are in a vacuum
chamber, at a pressure of about 3~$\times$~10$^{-6}$~mbar. The
vacuum is provided by a vibration-free ion pump. The whole
experiment lies inside a clean room (class 1000), with thermal
stabilization at the level of $\Delta T$ = 0.1$^\circ$, but showing a long term drift of the order of half a degree per day. The motion
of the resonator is detected by means of a Michelson type
interferometer set-up with a 1 mW amplitude stabilized He-Ne laser
at 633 nm. The interference signal is collected by a photodiode, whose 
voltage readout $V_M$ converts into distance by using:
\begin{equation}\label{eq:fringe}
    \delta x = \frac{\lambda}{2\pi}\frac{V_M}{V_{fr}}.
\end{equation}

where $\lambda=633 $ nm and $V_{fr}$ is the voltage corresponding to a interference fringe, in our case normally $V_{fr}=2.3$ V.

In order to achieve the maximum
sensitivity of the resonator, particular care has been taken to
minimize its free motion. The whole apparatus is mounted on a passive
low frequency damper and on a floating optical table. As explained
next, we want to minimize the low frequency noise, below a few
Hz: the passive damper (Minus-k Technologies 250 BM 6) is a
mechanical low pass filter  with a cut-off frequency at 0.5\,Hz in the
transmissibility curve. The clean room also represents a quiet
environment to avoid acoustical noise: all the electronics lie
in a separate control room. 
Moreover, a special metallic enclosure has been constructed around the apparatus. This enclosure has walls equipped with copper tubes through which temperature stabilized water flows, thus allowing very good long term stability of the entire system and thus eliminating the temperature drifts of the clean room. We discovered that temperature stability is crucial in order to perform repeated measurements over several days. In our case a stability of 0.1 degree of the experimental apparatus has been obtained for a time span of several days.

\subsection{Detection technique}

The set-up presented in this paper is based on an amplitude homodyne
detection technique, already described in \cite{bressi01a}. 

A static detection scheme is not sensitive enough, because it is
affected by low-frequency drifts due to thermal expansion,
electronic instabilities, drift of the laser frequency.
For this reason we use the dynamical homodyne detection technique: we
move the source with a periodical movement in the direction of the
resonator. The Casimir force acting between the two surfaces makes
the resonator move, and its motion is detected using the
interferometer. If the chosen frequency is well below the resonator
proper frequency, a fixed phase difference will be present between
source motion and resonator motion, thus allowing for vector
averaging of the interferometer signal.

If between the source and the resonator there is a spatially
dependent force $F(d)$, then the equation of motion of the
equivalent harmonic oscillator is
\begin{equation}\label{harmonic-oscillator}
    m\ddot{x}_r(t) = -m{\omega}_r^2x_r(t) + F(d),
\end{equation}
where $x_r(t)$ is the position of the resonator, $\omega_r = 2\pi
\nu_r$ is its angular proper frequency and $m$ its mass.

We can decompose the separation $d$ between source and resonator
into two components, namely a fixed distance $d_0$ and a
time-dependent component $x_s(t)$, due to the movement of the
source, that is $d = d_0 + x_s(t)$ ($|x_s/d_0| \ll 1$). The free motion of the resonator
is considered much smaller and is neglected. In the case of a
force of the
form $F(d) = C/d^n$, a periodic modulation of the position of the source $x_s = x_s^0
\cos \omega_s t$ (with $\omega_s\ll\omega_r$), results in (at first order):
\begin{equation}\label{eq:force}
    F(t) = \frac{C}{d_0^n}\left( 1 - n\frac{x_s^0}{d_0}\cos \omega_s t
    \right).
\end{equation}
The oscillation of the resonator  will be at the same frequency, and
its amplitude together with the value of the spring constant of the
resonator allows a calculation of the force parameters. In the Fourier
spectrum of the solution $x_r(t)$ of equation (\ref{harmonic-oscillator})
there is a peak at the frequency $\nu_s=\omega_s / 2 \pi$, whose
amplitude is:
\begin{equation}\label{eq:amplitude}
A_s^0 = n\frac{C x_s^0}{m\omega_r^2 d_0^{n+1}}.
\end{equation}

An overall calibration of the system is obtainable using
controllable electrostatic forces. In fact, a constant voltage $V$
between the resonator and the source will induce a force

\begin{equation}\label{eq:calibration01}
F_V(d) = -\frac{1}{2}\epsilon_0S\left(\frac{V}{d}\right)^2,
\end{equation}

It is then possible to compare the amplitude induced by the Casimir
force $A_C$ and the one obtained by the voltage calibration at fixed
bias $A_V$:

\begin{eqnarray}
A_C &=& \frac{1}{60} \frac{\pi^2\hbar c S}{m\omega_r^2 d_0^5} x_s^0  \label{eq:ac}\\
A_V &=& \frac{\epsilon_0 S V^2}{m\omega_r^2 d_0^3}x_s^0. \label{eq:av}
\end{eqnarray}

The calibrations are useful to infer the
parameters common to both expressions, without relying on their
direct determination.

\subsection{Stray effects}

In order to measure the Casimir force, one must be able to
eliminate or compensate other forces acting between the two plates. In
particular, forces of electrostatic nature can be present.
The origin of these forces can be divided into two different categories: residual bias voltage and patch potentials.

Even when two conductors are shortened together, a residual bias $V_0$ could be present between them due to contact potential. A counterbias $-V_0$ eliminates the effect of this residual potential, but one has to be sure that its value is not changing with time and gap separation. A careful study must be performed. Counterbias accuracy must be kept at a few mV level in order to disentangle the contribution of the Casimir force.

It is well known that local changes in the crystallographic directions exposed at the surface of a clean polycrystalline metal, as well as surface contaminations, give
rise to a point-dependent surface density $m(\vec{x})$ of dipoles on the surface of a metal \cite{speake}. The presence of such an inhomogeneous double layer of charges results in a spatially varying electrostatic potential $V (\vec{x})$ at the metal surface, relative to its interior, known as "patch" potential.
This zero-mean electrostatic component cannot be eliminated by a counterbias, and  gives rise to a force whose distance dependance is related to the spatial distribution of the patches. 
In order to calculate the force of attraction between the two plates, it is convenient to describe the patches with a characteristic length scale $\lambda_L$, which satisfies the condition:

\begin{equation}
d \ll \lambda_L \ll L
\end{equation}
where $d$ is the gap separation between the plates having lateral dimension $L$. The patch potentials $V(\vec{x})$ can be split as:
\begin{equation}
V(\vec{x}) = V^{(L)}(\vec{x}) + V^{(S)}(\vec{x})
\end{equation}
In this equation, $V^{(L)}(\vec{x})$ represents the {\it long wavelength} component, that does not vary appreciably under displacements smaller than $\lambda_L$ along the surface, while $V^{(S)}(\vec{x})$ represents the {\it short wavelength} component. Physically, $V^{(L)}(\vec{x})$ can be thought
to arise from the charges of a few isolated impurities distributed over the surface of the plates, while $V^{(S)}(\vec{x})$ describes the patch potentials of the microscopic crystallites
forming the surfaces. The corresponding variances $\sigma^2_L$ and $\sigma^2_S$ of the two components give rise to a force with a characteristic distance dependence. Following \cite{speake} it is possible to obtain a force of the form:

\begin{equation}
F(d) = \epsilon_0 L^2 \left( \frac{\sigma^2_L}{2d^2}+
\frac{2\sigma^2_S}{k^2_{\rm max}-k^2_{\rm min}}\int^{k_{\rm max}}_{k_{\rm min}} dk
\frac{k^3}{\sinh^2(kd)}\right) \label{eq:forza}
\end{equation}
where the wave numbers $k_{\rm max,min}=2\pi/\lambda_{\rm min,max}$ are related to the maximum and minimum sizes $\lambda_{\rm max,min}$ of the crystallites.

\section{Experimental results}\label{sect:exp-results}

The homodyne detection technique has been optimized looking
for  the modulation frequency of the source position providing the best signal to noise ratio. Long noise spectra are taken at the beginning of each measurement session, as a result 
the source is normally modulated at frequencies between 7 Hz and 14 Hz. This frequency range, well
below the resonator proper frequency, ensures a complete decoupling between source motion and resonator itself due to mechanical and electrical pick-ups, as it was checked by a very long run with the plates kept at a separation of about 50 microns. Most of the measurements have been performed with 45 nm amplitude for the source motion. The resonator motion noise was normally at the level of 
$3\cdot 10^{-11}$ m/$\sqrt{\rm Hz}$, but this figure was stable only at night, degrading by a factor of 2 during daytime. 

By imposing an external bias voltage $V_g$ to the gap, it  is then possible to calibrate the apparatus. This is done in order to obtain the value of the residual bias voltage $V_0$, the effective resonator mechanical response and the absolute gap width separation. Again, for a detailed description of these steps, we refer to a previous paper \cite{nostro}, while presenting here the most recent results. 

\begin{figure}[htb]
  \includegraphics[width=6in]{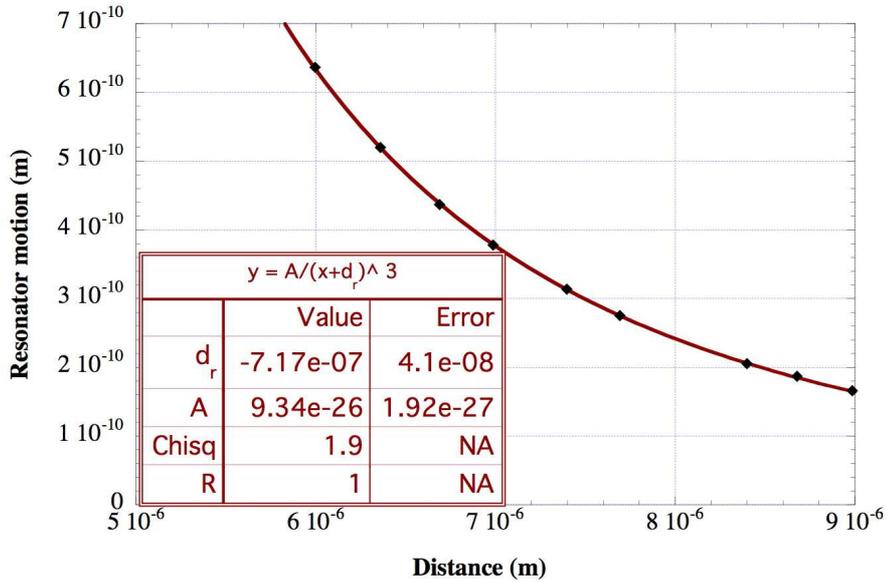}\\
  \caption{Measured signal versus distance when imposing an external  bias voltage in the gap. The solid line is the best fist with a $1/d^3$ function. The experimental error for each measurement is smaller than the marker size.}\label{superbias}
\end{figure}

By applying different values of an external bias voltage in the gap at fixed distance between the plates, one obtain the residual bias $V_0$. The resulting value was $V_0=244.7 \pm 0.8$ mV, with a stability of 1-2 mV over a few days period. For longer time scales changes up to 10 mV have been observed: the reason for these changes is not yet clear to us, a possibility is the temperature of the apparatus. The value of $V_0$ has also been measured for different distances, and no change has been observed.

The measurement of the resonator motion versus gap separation, while keeping the external bias $V_g$ constant, provides the reference distance $d_r$ and also  a cross check of the resonator mechanical response. The absolute gap separation is obtained by adding to the reference distance $d_r$ the displacement of the source given by the transducer PZT1. 

Figure \ref{superbias} shows a typical calibration response. The important parameter is the error on $d_0$, which is 40 nm. The coefficient $A$ differs from the expected one by about 5 \%.

\begin{figure}[htb]
  \includegraphics[width=6in]{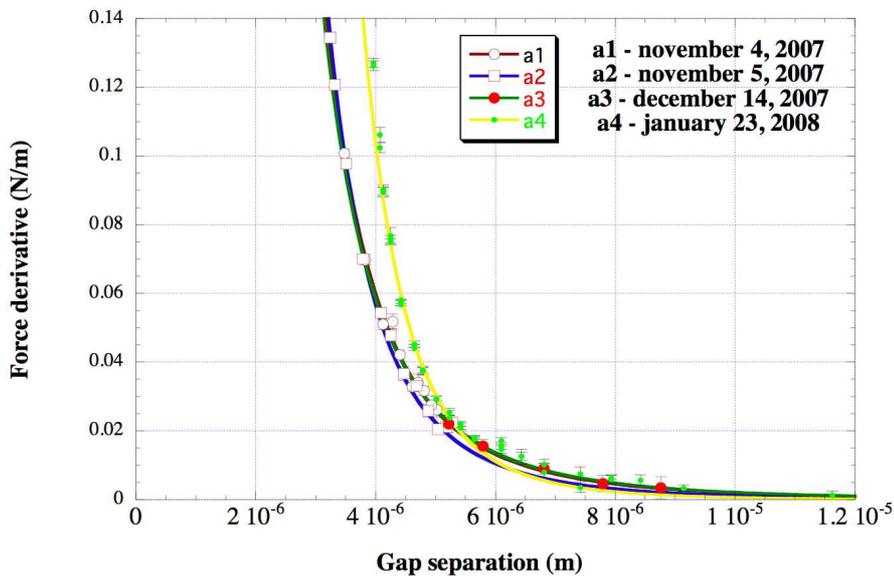}\\
  \caption{Summary of the measurements performed in the Casimir like configuration, i.e. a net zero voltage in the gap. Four different sets have been taken, corresponding to different measurement days. The abscissa is the absolute distance between the plates, as obtained by adding to the reference value $d_r$ the displacement of PZT1.}\label{Caslike}
\end{figure}

Casimir-type measurement are performed when $V_g=-V_0$, thus having a net zero voltage between the plates. Figure \ref{Caslike} shows the results obtained in 4 different measurement runs, performed in different days as indicated. The abscissa of the figure is the real gap separation already corrected using $d_r$. The ordinate shows the force derivative, i.e. the pulsating term of equation (\ref{eq:force}). Data are fitted with a function of the form $A/d^n$, where $A$ and $n$ are free parameters and $d$ is the gap separation. Results of the fits are rather astonishing: for $n$, values are in the range 3.5-5, and $A$ results to be way too large for being due to the Casimir force. Imposing a Casimir type dependence ($n=5$),  the experimental coefficient $A$ is about 100 times larger than expected.

It is then important to try to understand the reasons for these results. The presence of an uncompensated bias voltage is excluded since a too large value would be necessary, much larger than the error on the determination of $V_0$. The influences of AC voltages on the gap have been eliminated by using filters along the connecting cables and replacing the cables themselves. Also the effect of the residual pressure was studied by changing the amount of gas in the chamber: no variation has been seen. A more detailed analysis concerns the presence of patch effects.

\section{Patch effects}\label{sect:patch}

\begin{figure}[htb]
  \includegraphics[width=6in]{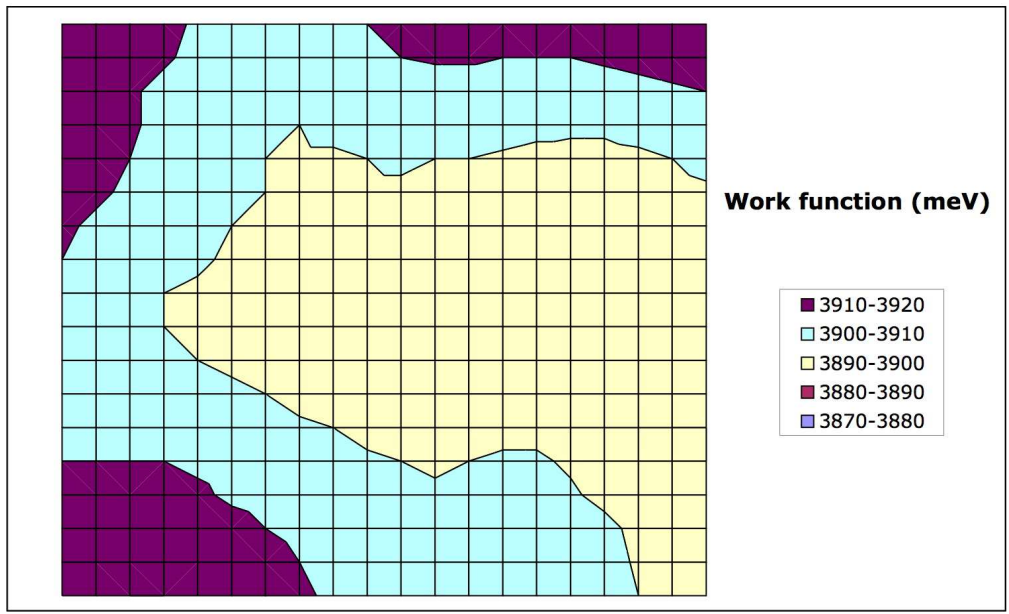}\\
  \caption{Kelvin probe analysis of a sample of bulk aluminium of the same type of the one used in the apparatus for measuring the Casimir effect. The measuring tip has a diameter of about 2 mm, each little square has a 0.35 mm side. }\label{Kelvinp}
\end{figure}

In order to study the presence of patches in the  surface of the plates, a Kelvin probe analysis \cite{Kelvin} has been done using plates similar to those mounted on the apparatus. Measurements of the work function across the samples were made by scanning the surfaces with a tip having an approximate diameter of 2 mm, with steps of 317.5 $\mu$m.
The data (see Figure \ref{Kelvinp}) showed that the work function exhibits significant fluctuations over distances of the
order of one millimeter, and therefore this  should be regarded as a long wavelength component of the patch potential in a force measurement involving separations  of the order of a few microns. The variances of the work functions were found to be in the range 10 - 30 mV.

Another analysis that we performed was the X ray diffractometry, in order to determine the grain size of the microscopic crystallites forming the surfaces. This analysis gave a 0.1 $\mu$m size for the grains forming the aluminium bulk, and a 0.03 $\mu$m size for the aluminium coating on the silicon resonator. This analysis gives no information on the amplitude and variance of the work function associated with the crystallites, which should form the short wavelength component of the patch related force.

\begin{figure}[htb]
  \includegraphics[width=6in]{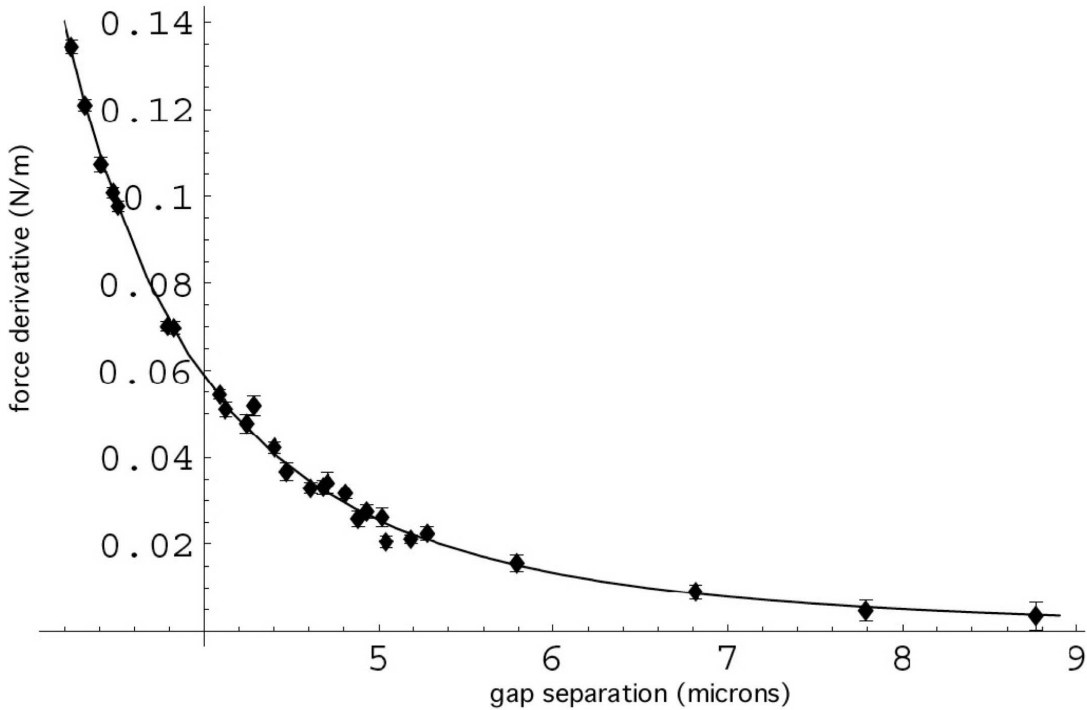}\\
  \caption{Fitting of the sets a1, a2, a3 of Figure \ref{Caslike}. The solid line is the best fit  with the derivative of the force (\ref{eq:forza}), for $L=1$ cm, $\lambda_{\rm min}=3\, \mu$m, $\sigma_S=91.3$ mV, $\sigma_L=51.6$ mV.}\label{patchf}
\end{figure}

The three sets of data a1, a2, a3 of Figure \ref{Caslike} have been fitted together with the derivative  $F'(d)$, with respect to $d$, of the force $F(d)$ of equation (\ref{eq:forza}), in order to extract from the fit the parameters related to the short and long wavelength components. The result is showed in Figure \ref{patchf}, the best fit is obtained with $\lambda_{\rm min}=3\, \mu$m, $\sigma_S=91.3$ mV, $\sigma_L=51.6$ mV. As it is clear from the figure, the fit is good, and the
probable values for $\sigma_S$ and $\sigma_L$ appear both very reasonable. The favored value for $\sigma_L$ is perhaps a bit larger
than what expected from the Kelvin probe measurement. However, it
should be considered that the tip used is very thick (2 mm), and therefore their
data actually represent smoothed measurements of the
work function, over areas of 4 mm$^2$. The smoothing
may have led to underestimate the real fluctuation of the
work function at scales of, say, 200-300 microns.

On the other side, the value $\lambda_{\rm min}=3\, \mu$m is indicating that the short wavelength component is not the one seen by the X ray diffractometry. There could be some other patches, for example again due to impurities, with micron size areas, that it is not possible to see with an independent surface analysis. This leaves still some open question on the understanding of our data.

 \section{Conclusions}
 
 We have set up an apparatus to measure the Casimir effect between parallel plates at separation in the range up to 5-6 microns. The possibility to achieve good parallelization and small gaps, down to 3 $\mu$m, using 1 cm$^2$ surfaces has been accomplished. The effort to measure the Casimir force is not yet successful, having encountered a much larger force of instrumental origin. Possible explanation of this force could be related to patch effects, but a clear identification of the patch potentials was not possible.
It has to be noted that the use of aluminium might have enhanced the presence of patches: for this reason another set up is under construction with gold plated surfaces, which should reduce the effect of patches. If this problem will persist also using gold, this will put a serious obstacle in the measurement of the Casimir effect at a large separation.

\section*{Acknowledgments}

We are grateful to the high skill of  E. Berto
(Universit\`{a} di Padova) and F. Calaon (INFN Padova). They helped
with the control of the vacuum set-up and the construction of mechanical
components. G. C. thanks his father for the help he provided to the
construction of the experimental set-up, tests and improvements
during the years. 


\end{document}